# Cascaded Logic Gates Based on High-Performance Ambipolar Dual-Gate WSe$_2$ Thin Film Transistors

AUTHOR NAMES

Xintong Li[†], Peng Zhou[‡], Xuan Hu[‡], Ethan Rivers[†], Kenji Watanabe[§], Takashi Taniguchi[§§], Deji Akinwande[†], Joseph S. Friedman[‡], and Jean Anne C. Incorvia[†]*

AUTHOR ADDRESS

[†]Department of Electrical and Computer Engineering, The University of Texas at Austin, Austin, Texas 78712, United States

[‡]Department of Electrical and Computer Engineering, The University of Texas at Dallas, Richardson, TX 75080-3021, United States

[§]Research Center for Electronic and Optical Materials, National Institute for Materials Science, 1-1 Namiki, Tsukuba 305-0044, Japan

[§§]Research Center for Materials Nanoarchitectonics, National Institute for Materials Science, 1-1 Namiki, Tsukuba 305-0044, Japan

*Corresponding Author, incorvia@austin.utexas.edu

ABSTRACT

Ambipolar dual-gate transistors based on two-dimensional (2D) materials, such as graphene, carbon nanotubes, black phosphorus, and certain transition metal dichalcogenides (TMDs), enable reconfigurable logic circuits with suppressed off-state current. These circuits achieve the same logical output as CMOS with fewer transistors and offer greater flexibility in design. The primary challenge lies in the cascadability and power consumption of these logic gates with static CMOS-like connections. In this article, high-performance ambipolar dual-gate transistors based on tungsten diselenide ($WSe_2$) are fabricated. A high on-off ratio of $10^8$ and $10^6$, a low off-state current of 100 to 300 fA, a negligible hysteresis, and an ideal subthreshold swing of 62 and 63 mV/dec are measured in the p- and n-type transport, respectively. For the first time, we demonstrate cascadable and cascaded logic gates using ambipolar TMD transistors with minimal static power consumption, including inverters, XOR, NAND, NOR, and buffers made by cascaded inverters. A thorough study of both the control gate and polarity gate behavior is conducted, which has previously been lacking. The noise margin of the logic gates is measured and analyzed. The large noise margin enables the implementation of $V_T$-drop circuits, a type of logic with reduced transistor number and simplified circuit design. Finally, the speed performance of the $V_T$-drop and other circuits built by dual-gate devices are qualitatively analyzed. This work lays the foundation for future developments in the field of ambipolar dual-gate TMD transistors, showing their potential for low-power, high-speed and more flexible logic circuits.

TEXT

Field effect transistors (FETs) based on 2D van der Waals materials have garnered significant interest in the post-Moore era due to their unique properties, such as high carrier mobility, ultra-thin thickness, tunable bandgap, and reduced short channel effects[1–3]. While most of the 2D thin film transistors (TFT) are unipolar, exhibiting either p-type or n-type conduction, ambipolar devices offer more flexibility in logic and analog circuit design by conducting both electrons and holes. This has been widely demonstrated in organic ambipolar transistors[4–6] and 2D material transistors[7–10]. Among the large atomically-thin material families, graphene, carbon nanotube (CNT), black phosphorus (BP), and certain TMDs like $WSe_2$ exhibit intrinsic ambipolar behavior without the need for doping. The primary challenge of ambipolar transistors is their higher off-state current compared to unipolar devices, particularly for small-bandgap materials such as graphene and BP[11,12]. Consequently, a dual-gate structure is advantageous in ambipolar devices, as it not only suppresses the off-state current, but also enables reconfigurable ambipolar TFTs that can be reversibly switched between p-type and n-type modes[12–15]. Furthermore, due to the independent input of the two gates, logic circuits made with ambipolar dual-gate TFTs can potentially achieve the desired operation with fewer transistors and lower power consumption than CMOS technology[16,17].

Several groups have previously demonstrated ambipolar TFT devices based on various 2D materials and their application in logic circuits. CNT-based devices demonstrate outstanding ambipolar behavior, but they face the challenge of achieving high current density due to their 1D nature, and the purification of semiconductive CNTs[17,18]. BP devices have shown promise in small supply voltage logic circuits[19], but they exhibit a relatively high off-state current, even with the dual-gate structure, due to the small bandgap of ~ 0.3 eV. The threshold voltage ($V_T$) of the demonstrated BP devices shifts from 0 V, leading to increased static power consumption. Additionally, BP devices have a well-known air-stability issue. In contrast, TMD materials such as $WSe_2$ have a thickness-dependent bandgap, high carrier mobility, and good stability, making them ideal for low power consumption, high speed logic gates. For instance, logic gates with conventional static CMOS connections have been demonstrated using polarity-controllable $WSe_2$ ambipolar dual-gate transistors[20,21]. However, the key challenge is that the input and output voltages are not in the same range. In another study, electrically tunable homojunction devices made of $WSe_2$ were used to achieve reconfigurable multifunctional logic circuits with fewer transistors using pass transistor logic, wherein both gate and drain nodes serve as inputs[22]. However, pass logic circuits have issues with finite input impedance and the accumulation of unsaturated output, necessitating the insertion of inverters or buffers between stages[16]. The conventional CMOS-like connection, in which transistors are connected to the supply voltages, is immune to those problems, and is shown to be superior in most cases with respect to speed, area, power dissipation, and power-delay products[23]. To date, logic gates based on TMD ambipolar transistors have not been made cascadable while maintaining a low power consumption, which is the basic requirement that the output and input voltages of each stage must be in the same voltage range, so that the output of the previous stage can be directly used as the input of the next stage. Moreover, the speed considerations of various gate structures in ambipolar dual-gate transistors have not been thoroughly analyzed.

Here, we outline the requirements for ambipolar dual-gate transistors to be used in low power consumption, high-speed, cascaded logic gates. Firstly, the transistor must exhibit a suitable $V_T$ for both p-type and n-type branches to enable cascadability and low power consumption in logic gates, a challenge that has persisted in previous works. If either the p- or n-mode operates in depletion mode, the transistor cannot be fully turned off using standard gate inputs. Instead, a voltage outside the operation voltage range must be applied to ensure accurate logic output and reduced off-state current[19,21]. Thus, an ideal ambipolar device should have a $V_T$ slightly larger than 0 V for n-type and slightly less than 0 V for p-type. Secondly, the transistor must possess good current-carrying ability, especially low contact resistance, for both types of transport in order to achieve high-speed applications[24,25]. In principle, metal-semiconductor contacts cannot be made Ohmic for both types of carrier injection simultaneously due to band alignment, and the contact of ambipolar devices generally forms a Schottky barrier. Therefore, proper contact engineering is necessary to maximize both electron and hole current[26,27]. Thirdly, the p-type and n-type current should be made as symmetric as possible to optimize the noise margin, which also requires suitable contacts. Finally, hysteresis control is essential to minimize $V_T$ shift during operation[28].

In this work, high-performance ambipolar dual-gate TFTs based on hexagonal boron nitride (hBN) sandwiched $WSe_2$ are fabricated, demonstrating a high on-off ratio for both p-type and n-type, an ideal subthreshold swing (SS), a proper $V_T$, and minimal hysteresis. For the first time, we show low power consumption cascaded logic gates based on TMD ambipolar TFTs with static CMOS connections, offering greater flexibility in circuit design with fewer transistors. Furthermore, the speed performance of the different gate structures in dual-gate TFTs are analyzed.

**Results and Discussion**

**Ambipolar dual-gate $WSe_2$ TFT fabrication and operation.** To meet the requirements for low power consumption cascaded logic gates, the device structure and contacts must be carefully engineered. Fig. 1a illustrates the $WSe_2$ TFT, in which a 4 nm thick $WSe_2$ flake is etched and sandwiched between two layers of 15 nm thick hBN. To facilitate hole and electron injection, 15 nm platinum and nickel are utilized as the bottom contact metal of the source and drain (SD) due to their relatively high and low work functions, respectively. Bottom contact is one of the widely used methods for providing damage-free contact between metal and 2D materials, as traditional top contact metal deposition such as chromium and titanium has been shown to penetrate and damage the 2D flakes, increasing contact resistance[29]. The control gate (CG) and polarity gate (PG) are positioned below and above the channel, respectively, enabling independent tuning of the electrostatic control of the two gates by altering the hBN thickness and gate metal. This is one of the crucial methods to ensure that input and output operate within the same voltage range for cascadability. The CG gates most of the channel area, functioning similarly to a conventional CMOS gate by controlling the carrier density of the channel. In contrast, The PG primarily gates the channel above the contact area, thus modulating the charge injection through the Schottky barrier. The CG thickness is also set at 15 nm, identical to the SD contacts thickness, to provide a flat surface for the $WSe_2$ flake. Fig. 1b displays an optical image of a typical TFT device with a channel length and width of 1.8 μm and 6 μm, respectively. Fig. 1c presents the symbol of the ambipolar dual-gate TFT used in this paper, which is a slightly modified version of the common dual-gate transistor symbol, to highlight the distinction between the PG and CG.

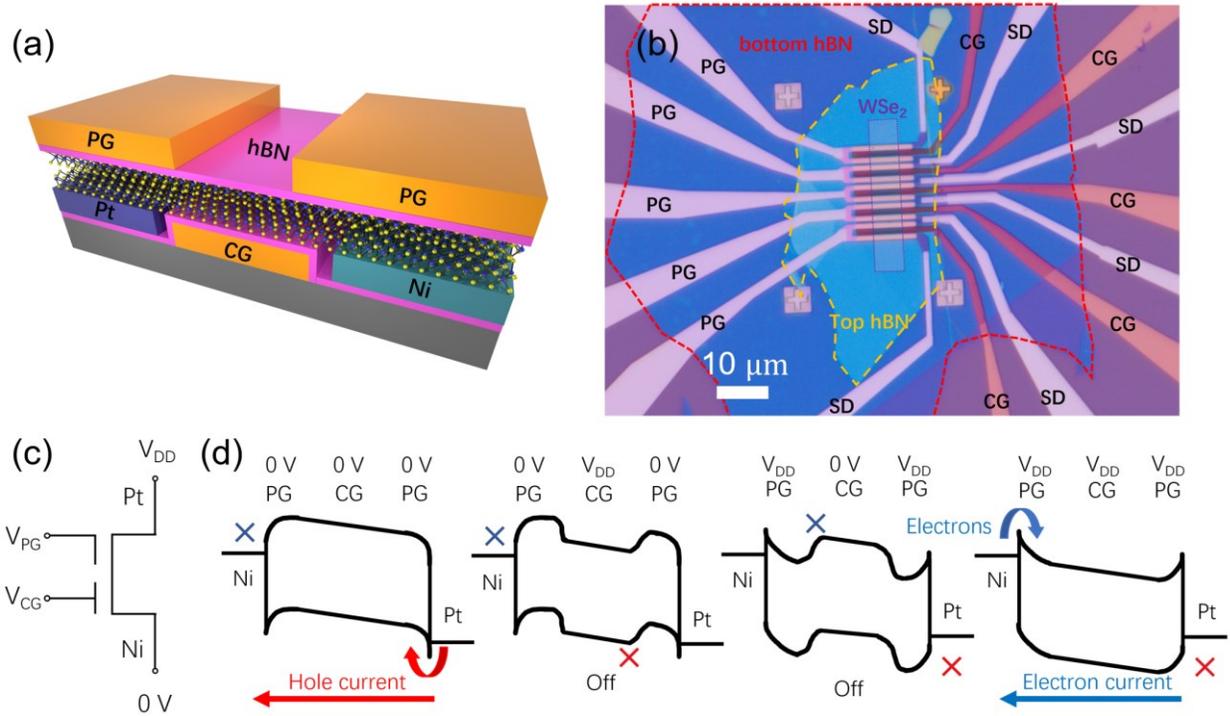

Figure 1. (a) Schematic representation of the ambipolar dual-gate WSe$_2$ TFT device structure, featuring a 4 nm-thick WSe2 flake etched and sandwiched between two layers of 15 nm-thick hBN. Source and drain (SD) contacts consist of 15 nm platinum and nickel, respectively (image not to scale). The control gate (CG) is embedded within the bottom hBN, regulating the carrier density in the channel. The polarity gate (PG), located atop the upper hBN layer, controls the carrier injection via the Schottky barrier at the contact. (b) Optical image of a typical TFT device, with a channel length and width of 1.8 $\mu m$ and 6 $\mu m$, respectively. SD, PG and CG are labeled, and the red and yellow lines outline the bottom and top hBN layers, respectively. The rectangle highlights the etched WSe$_2$ flake. The scale bar represents 10 $\mu m$. (c) Symbol for the ambipolar dual-gate TFT utilized in this study, depicting PG on top and CG on bottom. (d) Four-state operation of the TFT. From left to right: the on-state with hole current, the off-state with holes forbidden in channel, the off-state with electrons forbidden in channel, and the on-state with electron injection. Note the distinct positions of the Pt and Ni work functions relative to the WSe$_2$ band structure.

Fig. 1d illustrates the dual-gate control of the device. Platinum has a high work function of ~ 6.35 eV, close to the valance band maximum of WSe$_2$, while nickel has a lower work function of ~ 5.0 eV, closer to the conduction band minimum of WSe$_2$[30]. This difference is reflected in the illustrations. When hole (electron) current flows from the platinum to the nickel electrode, hole (electron) injection is facilitated by the high (low) work function of platinum (nickel). As a result, contact resistances for both electron and hole conduction are reduced[31]. Supporting Information

S2 presents the IV test results of a dummy sample with a platinum-only TFT, a nickel-only TFT, and a half platinum, half nickel TFT fabricated on the same WSe$_2$ flake. Clearly, the modification of the n-type and p-type transport is observed. It is well studied that gating the film at the contact area significantly modulates the contact resistance[32], as shown in the PG control in Fig. 1d. When a negative V$_{GS}$ is applied to the PG, hole injection through the platinum is allowed, while electron injection is prohibited. Then, if CG is also applied to a negative V$_{GS}$, excessive holes will be attracted to the channel to form a hole current. However, if a positive or zero V$_{GS}$ is applied to the CG, the current will be cut off since the channel lacks excessive holes. When a positive V$_{GS}$ is applied to the PG, the situation is similar, as shown in the last two figures in Fig. 1d. Consequently, the four resistance states follow the truth table of an Exclusive OR (XOR) gate, where the resistance is low only when the CG and PG have the same logic input. This ensures that the two gates are logically equal, which differs from some previous designs[22].

**IV characteristics of the TFT showing dual-gate control**. The IV characteristics of a single dual-gate TFT are measured at room temperature. The transfer characteristic curves of drain current I$_{DS}$ vs CG voltage V$_{CG}$ as a function of PG voltage V$_{PG}$ are displayed in Fig. 2a and 2b, with V$_{DS}$ set at 1 V. In Fig. 2a, note that the source is at the Pt electrode, so $V_S = 1\,V$. The curve reveals that when $V_{PG\_S} = V_{PG} - V_S = -1\,V\ to\ -4\,V$, the transistor operates purely as a high-performance p-type transistor, with a high on-off ratio exceeding 10$^8$, a low off-state current of ~ 100 fA, a V$_T$ of -0.6 V, and a steep SS of 62 mV/dec, which is close to the ideal value at room temperature. The red arrows indicate the sweep directions, and hysteresis is negligible, due to the hBN sandwiched structure that provides an atomically smooth and defect-free interface[33]. Note that the CG and PG voltages are not applied to the limit of the hBN to protect the device, since the safety voltage of hBN is ~ 0.4 V/nm[34]. The on-state current is expected to be significantly larger if a $V_{GS}$ of 6 V is applied. The highest two-port hole mobility achieved in our devices is ~ 40 $cm^2/(Vs)$, as shown in Supporting Information S3, which approaches the phonon-scattering mobility limit of few layer WSe$_2$ at room temperature. The flat region at low $V_{CG}$ results from the contact resistance limited by the PG being dominant in series. The n-type transport shown in Fig. 2b, on the other hand, is improved compared with previously reported ambipolar devices, but still has room for improvement. An on-off ratio larger than $10^6$, an off-state current of ~300 fA, a $V_T$ of ~ 0.4 V, and a SS of 63 mV/dec are achieved. Though the performance is sufficient for logic applications, the n-type on-state current is generally 2-10 times smaller than the p-type current. The primary cause for the n-type behavior is the higher Schottky barrier of the nickel-WSe$_2$ contact, as the work function of nickel is still larger than the electron affinity of WSe$_2$. A metal with a smaller work function could be employed as the contact in the future to further enhance the n-type behavior. However, the fabrication process must be optimized since most of these metals are prone to oxidation in air. The $I_D$-$V_D$ curves of the device can be found in Supporting Information S3.

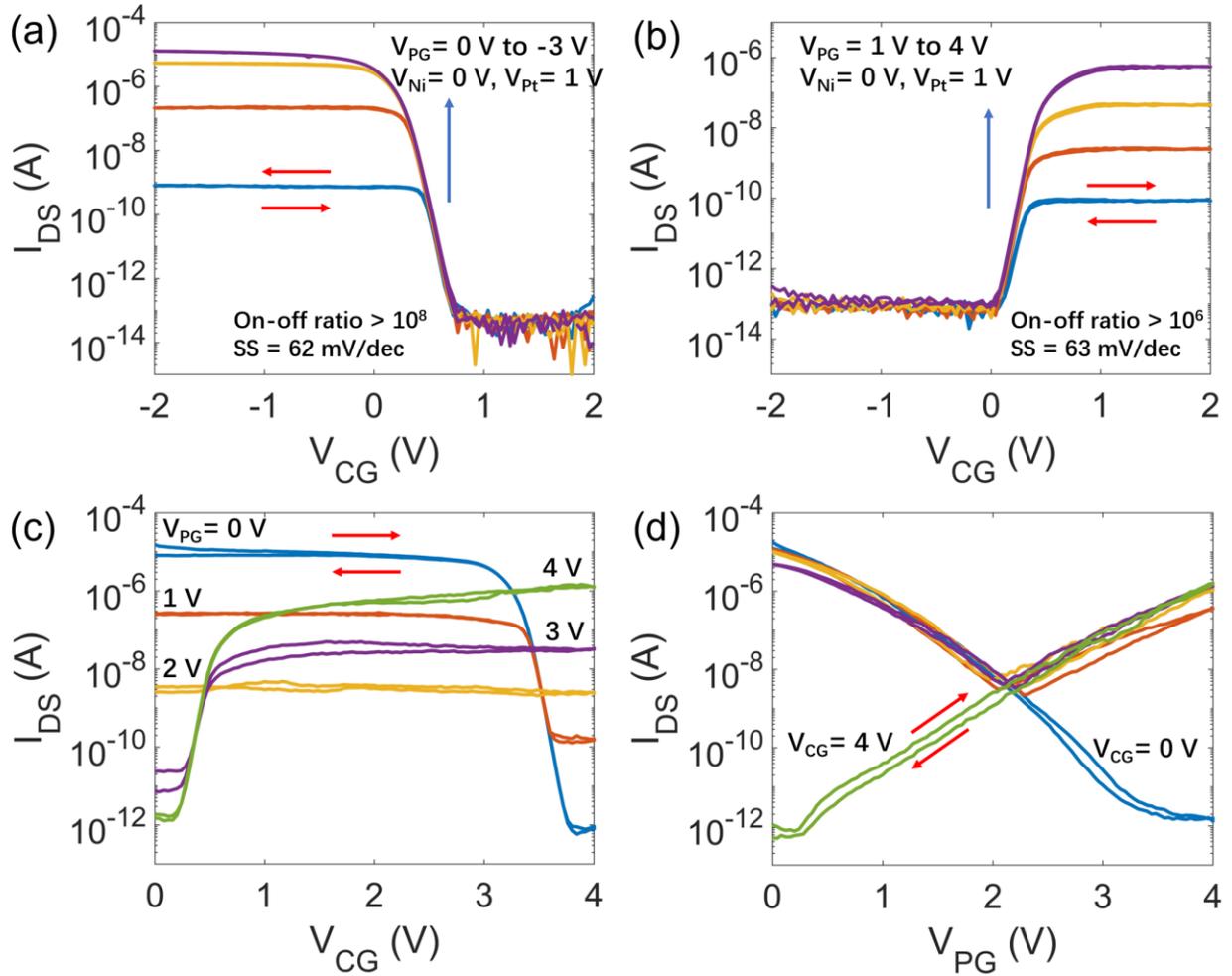

Figure 2. (a) Transfer characteristics of $I_D$ plotted against $V_{CG}$ for various negative $V_{PG}$ values, under a $V_{DS}$ of 1 V. Note that $V_S$ is 1 V for p-type TFT. The device achieves a high on-off ratio exceeding $10^8$, a low off-state current of ~ 100 fA, a $V_T$ of -0.6 V, and an ideal SS of 62 mV/dec for p-type. The sweep directions are indicated by red arrows, and negligible hysteresis is observed. (b) Transfer characteristics of $I_D$ versus $V_{CG}$ for various positive $V_{PG}$ values. $V_S$ is 0 V for n-type TFT. An on-off ratio exceeding $10^6$, an off-state current of ~ 300 fA, a $V_T$ of 0.4 V, and a steep SS of 63 mV/dec are achieved. (c) (d) Transfer characteristic curves under an operation voltage of 0-4 V, with $V_{DS}$ fixed at 4 V. Note that for n-type transport, $V_S$ = 0 V; for p-type transport, $V_S$ = 4 V. The device demonstrates correct four-state resistances with an off-state current of ~ 1 pA. CG and PG follow distinct mechanisms for current modulation. These characteristics ensure the cascadability of the logic gates.

It is clear that the $V_T$ requirements discussed in the previous section have been effectively met. This can be attributed to proper contact engineering, the near-zero charge neutral point of $WSe_2$, and the hBN sandwiched structure, all of which are critical factors in controlling the $V_T$ of TFT.

Of particular importance is the charge neutral point of WSe$_2$, which is close to 0 V due to the absence of intrinsic doping[35]. Most TMD materials, such as MoS$_2$, have intrinsic electron doping that result in negative n-type $V_T$ and work in depletion mode[1,36]. These materials cannot be effectively turned off at $V_{GS}$ = 0 V, making them less suitable for use in cascaded circuits. Similar reasons also account for the need to shift gate input voltages to turn off the transistors in previous works. Fig. 2a and 2b demonstrate that the TFT device can be reconfigured using the voltage of PG to form high-performance p-type or n-type transistors. If the polarity gate is connected to $V_{DD}$ or $V_{SS}$, these transistors work as polarity-reconfigurable transistors with ideal gate control.

To ensure the suitability of the devices for cascaded logic circuits, it is crucial to examine their performance under specified operation voltage, where the gate inputs, supply voltage, and outputs are all within the same range. In this study, a voltage range of 0 V to 4 V is used. It will be demonstrated later that the circuit also operates under smaller voltages. The transfer characteristics for CG and PG are shown in Fig. 2c and 2d as a function of the PG and CG voltages, respectively, with $V_{DS}$ fixed at 4 V to examine the off-state current in logic circuits, since the supply voltage always drops on the off-state transistors. The desired performance and resistance state are clearly maintained. The transistor is on with electron (hole) current when CG and PG are both at 4 V (0 V) and is off when CG and PG are at different logic input levels, with an off-state current of ~ 1 pA. The hysteresis resulting in a $V_T$ shift remains small for CG, while for PG it is slightly larger due to the inevitable lack of hBN substrate at the SD bottom contact area. The difference between the CG and PG transfer curves can be attributed to the distinct mechanisms of a conventional gate of a TFT and the gating behavior of a Schottky barrier FET (SB-FET). The former can be simply described by the long-channel model of a transistor in saturation region, while the PG gate follows the tunneling or thermionic emission in the Schottky barrier[37], where the current is modulated exponentially by the barrier height, which is reduced by increasing (decreasing) the PG voltage for n (p) type. Thus, the PG can tune the resistance in a wide range continuously with a slope lower than the CG.

**Cascadable logic gates made by WSe$_2$ ambipolar devices.** The performance of ambipolar dual-gate devices in various logic gates is then tested, starting with the inverter. Fig. 3a and 3b show the voltage transfer characteristics (VTC) of the inverter with the PG and CG as the inputs, respectively. While in most previous works, only inverters with CG input (the gate that only covers the channel region) are analyzed, we argue that examining the behavior of both PG and CG is necessary to interpret the characteristics of all subsequent logic gates, especially when the PG voltage is changeable to reconfigure the transistors, or is used as another input. The PG-input inverter is proven to work at $V_{DD}$ ranging from 4 V to 2 V, as shown in Fig. 3a. The gain is measured to be -26 at 4 V and decreases with $V_{DD}$. Due to the low off-state current of the ambipolar devices, the static power consumption is lower than 0.48 nW at $V_{DD} = 4$ V, dropping to 0.12 nW at $V_{DD} = 2$ V. The static current is slightly higher than the off-state current shown in Fig. 2, mainly due to variation between devices. The lowest static power consumption in our inverters falls below the noise level of the measurement tool and is less than 0.2 nW at $V_{DD} = 4$ V, much lower than most previous work. Surprisingly, despite the difference between the n-type and p-type on-state currents, the switching threshold $V_M$ and noise margin are almost immune to this imbalance in the PG-input configuration. The $V_M$ of the VTC, defined as the point where $V_{out} = V_{in}$, is 1.98 V for $V_{DD} = 4$ V. The noise margin low $NM_L$ is $0.47V_{DD}$, and the noise margin high $NM_H$ is $0.44V_{DD}$, which is significantly large. The current of the supply source is low even when the PG input approaches $V_M$. The current measurements and more about the inverters can be found in Supporting Information S4. The immunity of the noise margin to the p n imbalance and the low current are attributed to the distinct modulation of Schottky barrier of PG, shown in Fig. 2d. In contrast, the CG input configuration shown in Fig. 3b displays different behaviors, acting similarly to a conventional CMOS inverter. A higher gain of -48, and a static power consumption close to the PG-input inverter is measured. However, the $V_M$ and noise margins are strongly affected by the p- and n-branch currents. The $V_M$ for long-channel devices can be calculated by

$$V_M = \frac{V_{Tn} + r(V_{DD} + V_{Tp})}{1 + r} \qquad (1)$$

Here $r = \sqrt{\frac{k_p}{k_n}}$, $k_{p,n} = \frac{\mu_{p,n} C_{ox} W}{L}$, $\mu_{p,n}$ is the hole and electron mobility, $C_{ox}$ is the gate capacitance per unit area, $W$ and $L$ are the channel width and length, and $V_{Tn}$ and $V_{Tp}$ are the $V_T$ for n-type and p-type TFT, respectively. The measured $V_M$ in the CG-input inverter is 2.48 V, and $r$ is calculated to be 2.3, indicating a p- and n-type current ratio of ~ 5, which is consistent with the current measured in Fig. 2c and 2d. The $NM_L$ is $0.60V_{DD}$, and $NM_H$ is $0.34V_{DD}$. These values could be further improved by optimizing the n-type contact or adjusting the aspect ratio of the two transistors. When comparing the two configurations, there is a tradeoff between the $V_M$ and noise margin and the low power consumption with an input closer to $V_M$ in PG-input inverter, as well as the larger gain and steeper SS in CG-input inverter.

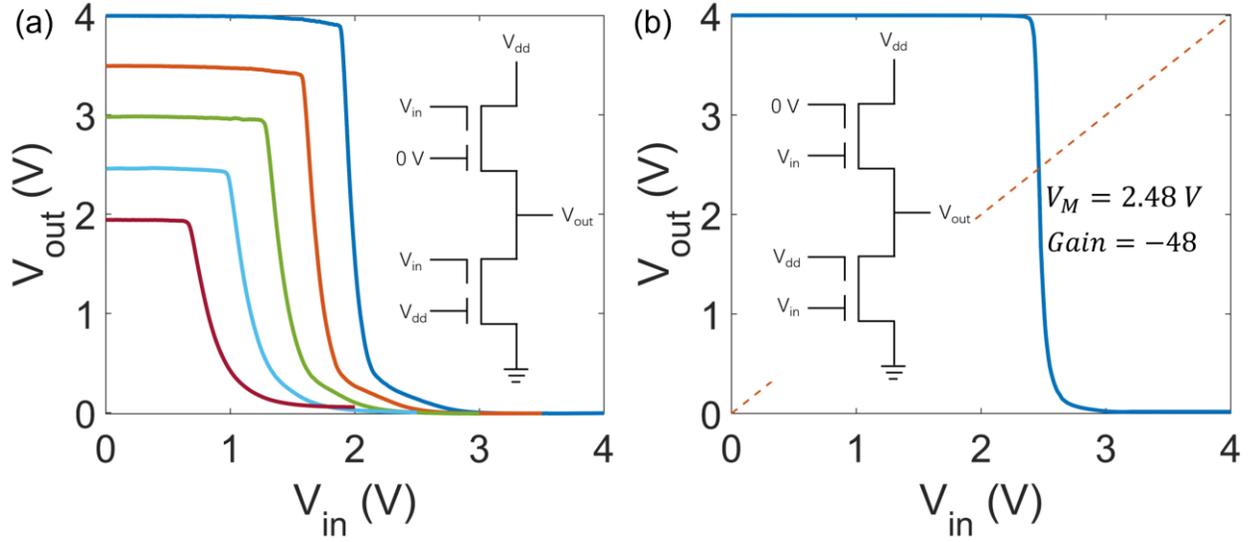

Figure 3. (a) The VTC of the inverter with the PG as input, operating under various $V_{DD}$ from 2 V to 4 V (red to blue). The measured gain is -26, with static power consumption < 0.48 nW, $V_M$ = 1.98 V, $NM_L = 0.47V_{DD}$, and $NM_H = 0.44V_{DD}$. The $V_M$ and noise margin show high immunity to the imbalance of n- and p-branch currents. (b) The VTC of the inverter with the CG as input under a $V_{DD}$ of 4 V, featuring a higher gain of -48. The $V_M$ and noise margin are affected by the imbalance of electron and hole currents. The cross point of the VTC with the red dotted line representing $V_{out} = V_{in}$ indicates a $V_M$ of 2.48 V.

Next, the ability to form various cascadable logic gates with standard pull-up network (PUN) and pull-down network (PDN) is demonstrated. Fig. 4a, b and c show schematics of the NAND, NOR and XOR gates made by the ambipolar dual-gate transistors. While the NAND and NOR gates act similarly to the CMOS circuits, the XOR gate benefits from the independent and symmetric input of the CG and PG. As a result, the transistor count is reduced from 8 to 4, the same as NAND and NOR gates. This allows for more flexibility in logic design with a reduced number of transistors and reduced power consumption. To demonstrate the cascadability, a buffer consisting of two cascaded inverters is schematically shown in Fig. 4d and tested. The corresponding timing diagram of the input and output of the logic gates and cascaded buffer is measured in Fig. 4e. Note that all the input, supply and output are in the same operation range of 0-4 V. Thus, a fully cascaded logic circuit based on 2D TMD ambipolar TFTs is demonstrated for the first time, with low static power consumption perfectly maintained, due to the $V_T$ control of the devices. Output voltage saturation is well achieved with negligible shift from 0 V or $V_{DD}$. The slight distortion of the shape of some output curves and a prolonged rise/fall time of ~20 ms are due to the large parasitic capacitance of the measurement setup, which is estimated to be approximately six orders of magnitude higher than the intrinsic capacitance from the device. For NAND and NOR gates, the output is nearly identical if the input A and B are applied to the PG. More information about the VTC of the buffer can be found in Supporting Information S5.

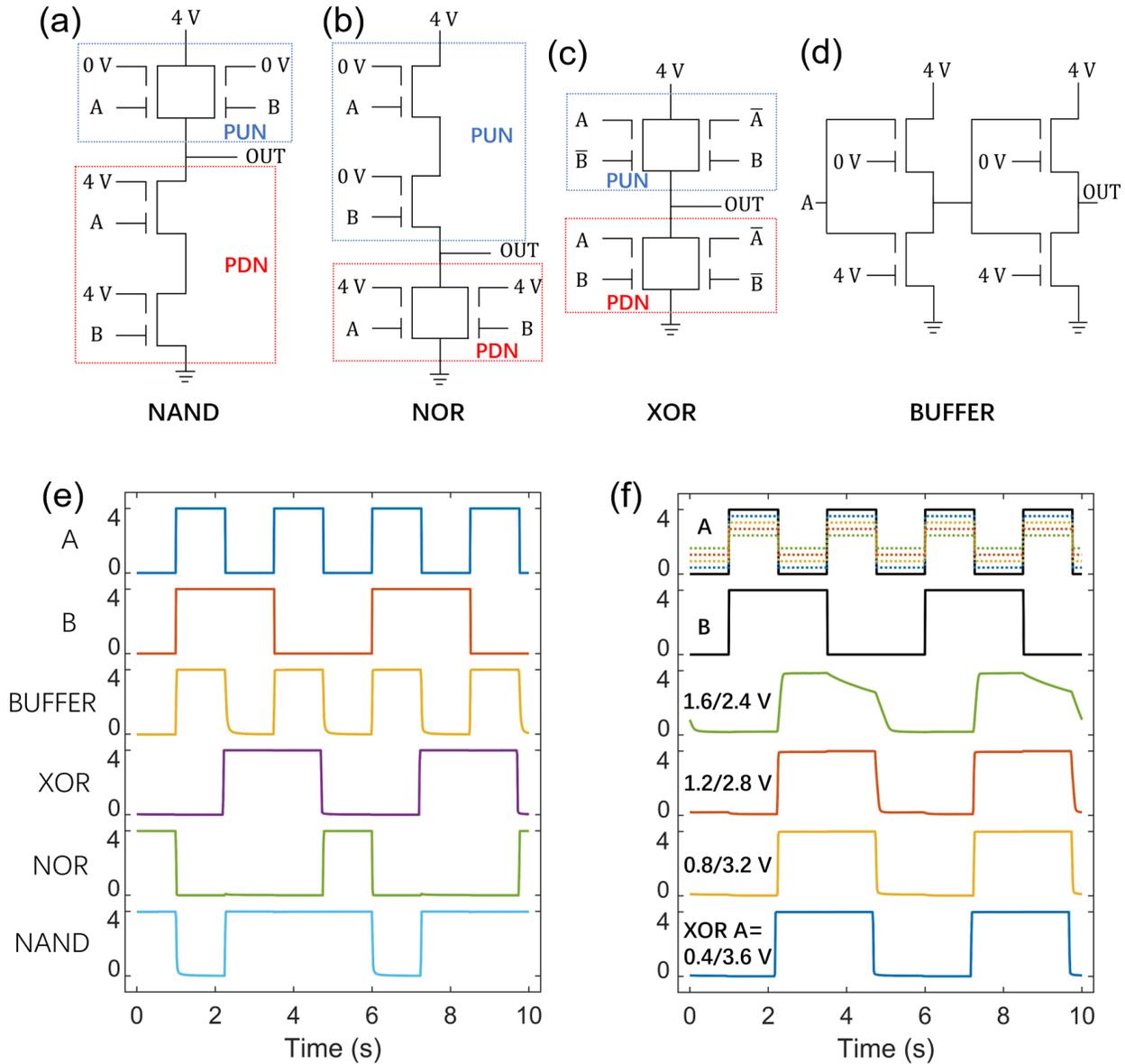

Figure 4. Schematics of (a) NAND, (b) NOR and (c) XOR gates constructed using ambipolar dual-gate transistors. PUN and PDN are circled. Both CG and PG gates serve as inputs for the XOR gate. (d) A buffer constructed using two cascaded inverters. (e) Timing diagram of input and measured output signals. Output voltage saturation is effectively achieved, with negligible deviation from 0 V or $V_{DD}$. Note that all inputs, supply, and outputs operate within the same operation voltage range, while static power consumption remains ideally low due to the $V_T$ control of the devices. The buffer demonstrates the cascadability of the logic gates. (f) Output curves of a 4-transistor XOR gate with the PG-input (A) high- and low-logic voltages deviated from 0 V and $V_{DD}$ by 0 V to 1.6 V. Correct output logic is maintained for input deviation up to ~ 1.6 V.

The demonstrated logic gates exhibit outstanding immunity to input noise or voltage shift, with different behaviors observed for CG and PG inputs. Fig. 4f shows the output curves of a 4-transistor XOR gate with the PG input (signal A) high and low logic voltages shifted from 0 V and $V_{DD}$ by 0 V to 1.6 V. The correct output logic is maintained up to an input noise of ~ 1.6 V, indicating a substantial noise margin. The rise/fall time increases with higher input voltage shift due to the decreased on-state conductance. Supporting Information S6 provides similar plots for CG input noises, with the fall/rise time not heavily affected by the CG input shift. Due to the smaller $NM_H$ of the CG-input inverters without aspect ratio adjustment, the gate can maintain the correct output with a high-logic CG input noise of < 0.8 V, which can be further improved using the method mentioned above for CG-input inverters. Remarkably, the static power did not increase significantly with PG input voltage shift up to 1.2 V, attributed to the PG control similar to that in the PG-input inverters. This property enables a special type of application, called "$V_T$-drop" logic gates, which offers more flexibility and even fewer transistors at the cost of output saturation and speed[38]. The following section discusses these topics and speed considerations.

**$V_T$-drop logic gates and qualitative speed analysis of the ambipolar dual-gate TFTs.** The high noise margin of the PG, the potential high noise margin of the CG through aspect ratio tuning, and the fact that the static current did not increase significantly with increasing PG input noises, make "$V_T$-drop" logic gates possible[38]. Fig. 5a, b and c illustrate several $V_T$-drop circuits, including a 2-TFT XOR gate, reconfigurable NAND/OR gate, and reconfigurable NOR/AND gate. The 2-TFT XOR gate is essentially half of the 4-TFT XOR gate (note that the A and B inputs are switched here for convenience of the measurements, but have no effect on the outputs. The timing diagram of A and B is the same as Fig. 4). The circuits in Fig. 5b and 5c are the same NAND and NOR gates, but the gate that was connected to ground or $V_{DD}$ is replaced with a third input C. Thus, when C is in logic low, the gate functions as a NAND and a NOR gate in Fig. 4. When C is in logic high, however, they become OR and AND gates, respectively.

In conventional CMOS gates, using a p-type transistor in the PDN or n-type transistor in the PUN is pointless because these transistors are turned off before the output fully reaches logic low or logic high, since the $V_{GS}$ is no longer a fixed value of $V_G - V_{SS}$ or $V_G - V_{DD}$, but is $V_G - V_{out}$. As a result, these designs do not have a full-swing, well-saturated output, but instead have an output with a voltage drop from $V_{SS}$ or $V_{DD}$. However, for ambipolar dual-gate devices, they can be reversibly switched between p-type and n-type. Allowing p-type transistors in PDN or n-type transistors in PUN can provide more flexibility and further reduce the number of transistors. For the 2-TFT XOR gate, the PUN becomes a single n-type transistor when both CG and PG are in high. In steady state, this transistor requires a $V_{DS}$ voltage close, but usually smaller than $V_T$, to partially turn on the transistor so that its current equals the off-state current in the PDN. This is why it is called a "$V_T$-drop gate". The OR gate and AND gate follow similar principles. The output of a 4-TFT XOR gate, a 2-TFT XOR gate, and a 2-TFT XOR gate with a $V_T$-dropped PG input are depicted in Fig. 5d. While the output of the 2-TFT XOR gates is at the correct logic level, it experiences a $V_T$-drop of 0.3 to 0.4 V from the $V_{DD}$ or ground when the PUN has a gate input of (4,4), or when the PDN is (0,0). The top yellow curve represents the situation where a $V_T$-dropped output is cascaded into the input of a second $V_T$-drop stage, indicating that the $V_T$-drop can accumulate between $V_T$-drop stages. Therefore, in order to pull up or down the voltage to $V_{DD}$ or 0 V, it is critical to carefully design the logic signal flow or to include non-$V_T$-drop logic gates such as those shown in Fig. 3 and 4 between several $V_T$-drop stages. Examples of output curves of NAND/OR and NOR/AND gates can be found in Supporting Information S7, and the estimated value of the $V_T$ drop can be found in Supporting Information S8. Due to the large noise margin and the immunity of off-state current to the input voltage drop of PG, the $V_T$-dropped outputs should preferably be sent to the PG of the next stage. This allows for more flexibility, a reduced number of TFTs, and lower power consumption, but comes with a tradeoff of the need for an increased $V_{DD}$ and lower speed. An example of $V_T$-drop circuit can be found in Supporting Information S9, where a full-adder with 12 transistors is illustrated, which is less than half of the 28 transistors required for CMOS-like logic designs.

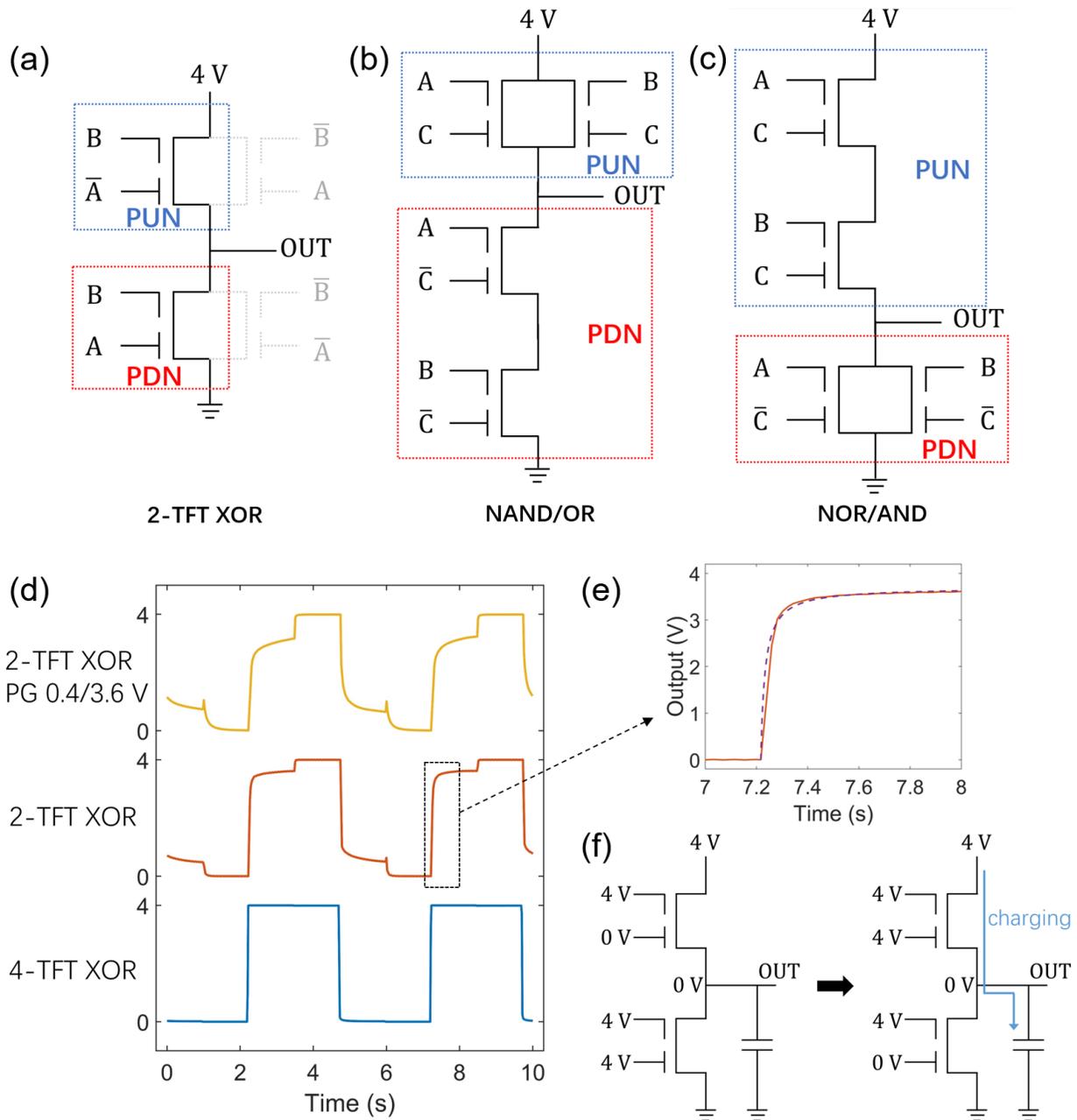

Figure 5. Schematics of several $V_T$-drop logic gates based on ambipolar dual-gate TFTs, including (a) a 2-TFT XOR gate, (b) a reconfigurable NAND/OR gate, and (c) a reconfigurable NOR/AND gate. These gates achieve more complex functions with fewer transistors, with a tradeoff of $V_T$-drop compared to full-swing output circuits. (d) The output of a full-swing 4-TFT XOR gate, a $V_T$-drop 2-TFT XOR gate, and a 2-TFT XOR gate with a $V_T$-dropped PG input. The $V_T$-drop states occur due to the presence of p-type transistor in PDN, or n-type transistor in PUN. Output logic is correctly maintained by the $V_T$-drop circuits. The $V_T$-drop output of the previous stage can also be used as the input (preferably PG) of the next $V_T$-drop stage, with an accumulated $V_T$-drop. Thus, non-$V_T$-drop gates must be used between several $V_T$-drop gates to pull up or down the voltage to VDD or 0 V. (e) Magnified view and fitted curve of the rising edge of a $V_T$-drop state, where the dotted line represents the fitted curve using equation (2), and the red line represents the measured data. The rise time is prolonged due to the charging/discharging through the diode-connected n-

type transistor in the PUN. (f) The schematic of the charging of the parasitic capacitor through the diode-connected n-type transistor.

Considering the speed performance of the $V_T$-drop gates, the fall/raise edges are distorted when the output jumps from low to $V_{DD} - V_T$, or from high to $V_T$, as shown in Fig. 5e, which is a magnified view of a single rising edge (Note that the precise measurements of fall/rise time or propagation delay in this study are limited by the parasitic capacitances presented in the measurement and interconnected systems, thereby allowing only qualitative analysis of speed performance). As the output initially rises quickly, the shape becomes distorted into a curve, and the rising gradually slows down. The rise time is ~ 60 ms for 75%, and significantly longer for 90%, which is considerably slower than the typical rise/fall time of non-$V_T$-drop gates, which ranges from ~ 10 to 20 ms (the large parasitic capacitance is assumed to be equal). This behavior can be explained by the schematic presented in Fig. 5f. Initially, the gate is in a non-$V_T$-drop state with a 0 V output, and the parasitic capacitor is fully discharged, as depicted in the left figure. Then when the CG input is reversed with a sharp fall/rise time, the instantaneous output remains at 0 V, as shown in the right figure. Consequently, the capacitor must be charged through the PUN to increase the output voltage. Here the PUN transistor is essentially diode-connected, and the charging of the capacitor versus time $t$ can be described by

$$V_{out} = (V_{DD} - V_{Tn})(1 - \frac{1}{\frac{k_n(V_{DD} - V_{Tn})}{2C}t + 1}) \qquad (2)$$

Here $C$ is the capacitor, $t$ is time starting from the start of the rise edge. The output voltage thus follows an inversely proportional curve and approaches $(V_{DD} - V_{Tn})$ when $t \to \infty$. This equation is used to fit the measured rising edge and the fitted curve is plotted in Fig. 5e. This diode-connected charging/discharging process is widely employed in other applications, such as the internal $V_T$ compensation of active matrix (AM) designs for OLED displays, where the $V_T$ value is programmed and stored in the capacitor within a few tens of $\mu s$ [39,40]. When the operation frequency increases, the charging/discharging of the $V_T$-drop circuits may be incomplete. Consequently, the actual output voltage drop might exceed $V_T$, and this speed tradeoff should be considered based on the specific application.

Similar analysis can be applied to non-$V_T$-drop logic gates and nearly all previous works involving dual-gate devices. The PG or any gates controlling the Schottky barrier injection operates differently from the CG, which solely controls the channel, as demonstrated in Fig. 2c and 2d. In cascaded circuits, the input fall/rise time can be as substantial as the output. The higher SS of PG input subsequently extends the fall/rise time and propagation delay of the output when compared to the CG input. It should be noted that the SS of PG can be further improved by thinning the PG dielectric or utilizing high-$\kappa$ 2D dielectric layers[37]. On the other hand, in terms of input noise, the PG input offers greater noise immunity while simultaneously maintaining a low static power consumption.

**Conclusions**

In this study, we outline the requirements for employing ambipolar dual-gate 2D TFTs in cascaded logic circuits, with $V_T$ control being one of the most critical aspects for cascadability and power saving. High-performance ambipolar dual-gate TFTs based on hBN-sandwiched $WSe_2$ are fabricated and tested. The transistors achieve a high on-off ratio greater than $10^8$, an off-state current of 100 fA, and an SS of 62 mV/dec for p type, as well as an on-off ratio greater than $10^6$, an off-state current of 300 fA, and an SS of 63 mV/dec for n type. By engineering the contact, dielectric environment, and gating, the devices are proven capable of forming cascaded logic circuits. PG-input and CG-input inverter configurations are tested, achieving gains of -26 and -48, along with exceptionally low static power consumption. Subsequently, standard XOR, NAND and NOR gates, as well as buffers made by two cascaded inverters with static CMOS-like connections are demonstrated, and the noise margins of these circuits are thoroughly analyzed. Futhermore, $V_T$-drop circuits enabled by the large noise margin and low power consumption of the devices are introduced and tested. The tradeoff between the number of transistors, power consumption, design flexibility, and output swing and speed is analyzed in detail. The future development direction of ambipolar $WSe_2$ TFTs is clear. Optimizing n-type currrent, particularly the n-type contact, is necessary. Improving the PG control over the Schottky barrier can be achieved by reducing the hBN thickness or using a high-$\kappa$ 2D insulator. Additionally, device scaling is also required to observe the behavior of short-channel devices[41]. In conclusion, this work demonstrates for the first time low-power consumption cascadable and cascaded logic gates based on ambipolar TMD TFTs, enhancing the understanding of the relevant devices and performance. This constitutes a critical step towards the application of such devices in logic circuits.

**Methods**

**Ambipolar dual-gate $WSe_2$ TFT fabrication.** A Si substrate with 285 nm $SiO_2$ is cleaned. Metal pads and course/fine alignment markers made from 80 nm of Cr/Au are defined using e-beam lithography (EBL) and e-beam evaporation. A 15 nm Cr/Au CG is patterned and deposited. The sample undergoes ozone ultraviolet (UV) cleaning before a 15 nm hBN flake is exfoliated from bulk and transferred onto the CG using a standard PPC/PDMS dry transfer technique inside an Ar-filled glovebox. A 200 °C high vacuum annealing is applied to improve hBN adhesion and eliminate tape residue and air bubbles. 15 nm Ti/Pd/Pt and 15 nm Ti/Ni contacts are then patterned and deposited, respectively.

A 4 nm $WSe_2$ flake is exfoliated from bulk onto another ozone UV-cleaned substrate. The sample is annealed at 200 °C before the flake is patterned and etched with $CHF_3$ and $O_2$ thermal plasma. The target sample, $WSe_2$ sample, and the sample with another 15 nm hBN are all annealed to remove absorbed $O_2$ and water. The samples are then placed into the glovebox immediately, and a dry transfer with PPC/PDMS is used to pick up hBN and $WSe_2$ and place them onto the contact area of the transistors. A 200 °C high-vacuum annealing is performed again to eliminate the residue and air bubbles. Finally, 30 nm of Cr/Au is patterned and deposited as the PG.

**Electrical testing of single devices and logic circuits.** Single device characterization is performed on a Cascade probe station with a Keysight 4156 semiconductor analyzer. The logic circuits are tested in a custom-made probe station. The devices are connected to form logic circuits by connecting the probes or by wire bonding on a printed circuit board (PCB). The measurement tools include Keithley 2600B, 2400 and 2401 sourcemeters and Keysight 33500B waveform generators.

ASSOCIATED CONTENT

The supporting information is available free of charge on the ACS publications website.

Table of recent experimental work on ambipolar reconfigurable logic gates based on 2D materials; IV characteristics of a platinum-only TFT, a nickel-only TFT, and a half platinum, half nickel TFT; Additional IV characteristics of the ambipolar dual-gate devices; Current measurement and more about inverters; VTC of the two-inverter buffer; Additional output of the logic gates with CG input voltage noise or deviation; Example of output curves of NAND/OR gates; Explanation of the $V_T$-drop; Schematic of a full adder with 12 transistors.

AUTHOR INFORMATION

**Corresponding Author**

*Email: incorvia@austin.utexas.edu

**Author Contributions**

The manuscript was written through contributions of all authors. All authors have given approval to the final version of the manuscript. X. L. fabricated the devices, carried out the measurements, and wrote the manuscript. P. Z. and X. H. assisted with the measurements. E. R. assisted with the fabrication. K. W. and T. T. provided the high-quality hBN bulk material. J. S. F. and D. A. supervised. J. A. I. conceived the project, led supervising the work, and wrote the manuscript.

**Funding Sources**

The authors acknowledge funding support from the National Science Foundation Electrical, Communications and Cyber Systems (NSF ECCS) Grant Number 2154285 (X. L., J. A. I) as well as funding from the NSF ECCS Grant Number 2154314 (P. Z., X. H., J. S. F.). D.A acknowledges the support of the Air Force Research Laboratory (AFRL) award FA9550-21-1-0460. E. R. acknowledges support from the NSF Center for Dynamics and Control of Materials: an NSF MRSEC research experience for undergraduates, under Cooperative Agreement No. DMR-1720595. This work was performed in part at the University of Texas Microelectronics Research Center, a member of the National Nanotechnology Coordinated Infrastructure (NNCI), which is supported by the National Science Foundation (Grant ECCS-2025227). The authors acknowledge the use of shared research facilities supported in part by the Texas Materials Institute and the Texas Nanofabrication Facility supported by NSF Grant No. NNCI-1542159. K.W. and T.T. acknowledge support from the JSPS KAKENHI (Grant Numbers 20H00354, 21H05233 and 23H02052) and World Premier International Research Center Initiative (WPI), MEXT, Japan.

**Notes**

The authors declare no competing financial interest.

ACKNOWLEDGMENT

We would like to thank Dr. Jaesuk Kwon for his advice on the electrical measurements. We thank the REU program in UT Austin supporting E. R. from the Center for Dynamics and Control of Materials: an NSF MRSEC.